\documentclass[aps,prb,twocolumn,floats]{revtex4}

\usepackage{amssymb}
\usepackage{amsmath}
\usepackage{graphicx}
\usepackage{bm}
\usepackage{mdframed}
\usepackage[pangram]{blindtext}

\begin{document}

\bibliographystyle{prsty}
\author{H\'{e}ctor Ochoa}
\thanks{These authors equally contributed to this work}
\author{Ricardo Zarzuela}
\thanks{These authors equally contributed to this work}
\author{Yaroslav Tserkovnyak}
\affiliation{Department of Physics and Astronomy, University of California, Los Angeles, California 90095, USA}

\begin{abstract}
Spin superfluidity, i.e., coherent spin transport mediated by topologically stable textures, is limited by parasitic anisotropies rooted in relativistic interactions and spatial inhomogeneities. Since structural disorder in amorphous magnets can average out the effect of these undesired couplings, we propose this class of materials as platforms for superfluid spin transport. We establish nonlinear equations describing the hydrodynamics of spin in insulating amorphous magnets, where the currents are defined in terms of coherent rotations of a noncollinear texture. Our theory includes dissipation and nonequilibrium torques at the interface with metallic reservoirs. This framework allows us to determine different regimes of coherent dynamics and their salient features in nonlocal magneto-transport measurements. Our work paves the way for future studies on macroscopic spin dynamics in materials with frustrated interactions.
\end{abstract}

\title{Spin hydrodynamics in amorphous magnets}

\maketitle

\section{Introduction}

The idea of low-dissipation, topologically-protected spin transport emphasized here does not rely on particle-like degrees of freedom, like in conventional mass/charge superfluidity, but on the orientational dynamics ascribed to some form of magnetic order.\cite{Sonin} Systems where both phenomena coexist are, for example, superfluid $^3$He, \cite{He3} and Bose-Einstein condensates of alkali atoms.\cite{rev} Recent advances in spintronics allow for the generation and detection of spin supercurrents in solid-state systems. The proposed platforms consist of electrically insulating easy-plane (anti-)ferromagnets,\cite{konig,so1,Chen,so3,afm} including the canted antiferromagnetic phase in the $\nu=0$ quantum Hall state of graphene, \cite{so4} which has motivated some recent experimental progress.\cite{exp_graphene1} A metastable spiraling texture hosts the spin superfluid, whose dynamics is triggered by spin-orbit torques at an interface and subsequently detected via the reciprocal pumping effects. 

Akin to conventional superfluids, the stability relies on the U(1) symmetry of the functional governing the macroscopic dynamics. This symmetry, however, breaks down in the presence of planar anisotropies. Collective macroscopic transport is only possible beyond certain current threshold, as long as the strength of the easy-plane anisotropy exceeds that of these detrimental perturbations.\cite{konig} In that regard, the thermal dependence of long-ranged drag signals recently reported in Cr$_{2}$O$_{3}$ is indicative of some form of coherent spin dynamics;\cite{Cr2O3} however, this effect is only observed in the second harmonic, suggesting that interfacial spin-transfer torques are ineffective in making the spin texture precess, possibly due to the presence of parasitic anisotropies near the interface.
\begin{figure}[t!]
\begin{center}
\includegraphics[width=0.9\columnwidth]{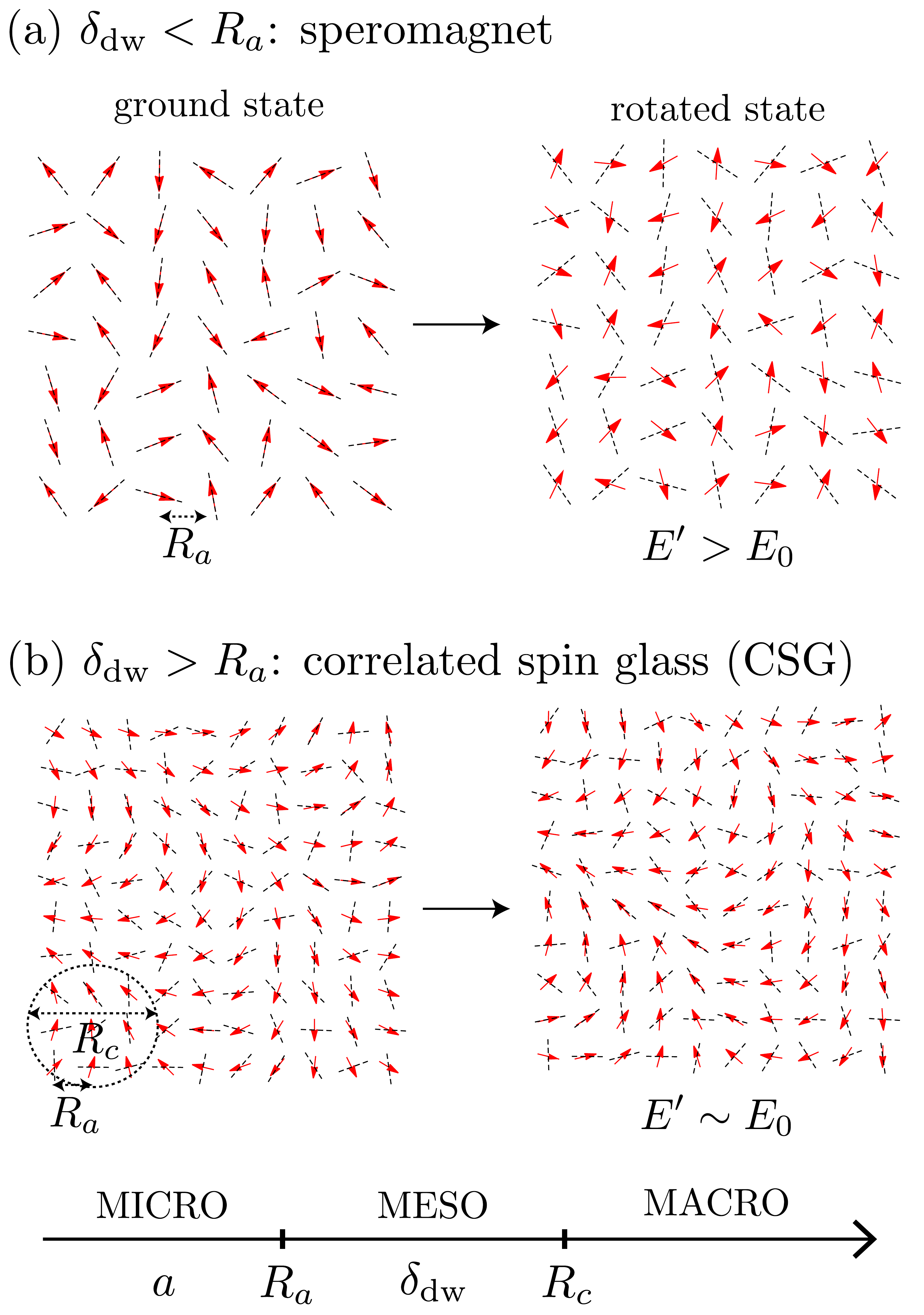}
\caption{Spin textures below $T_f$. (a) $\delta_{\textrm{dw}}<R_a$: the spins (red arrows) remain pinned to the anisotropy axes (dashed lines) defined by the local atomic arrangement. Collective spin rotations cost energy. (b) $\delta_{\textrm{dw}}>R_a$: exchange interactions stabilize a smooth spin texture on the scale of the grain size. Collective spin rotations correspond to soft modes. The hierarchy of length scales in this phase is shown at the bottom.}
\vspace{-0.5cm} 
\label{Fig1}
\end{center}
\end{figure}

In this article, we exploit the fact that structural disorder present in amorphous or polycrystalline materials can eventually frustrate these parasitic anisotropies in the exchange-dominated limit for magnetic interactions. For example, strong exchange interactions have been invoked to explain recent nonlocal transport measurements in amorphous yttrium iron garnet (YIG).\cite{exp} We consider, in particular, noncollinear spin textures below the freezing temperature that are smooth on the (microscopic) scale of the grain size. This is the 
so-called \emph{correlated spin glass} (CSG) phase, which is schematically depicted in Fig. \ref{Fig1}(b).\cite{CSG0,CSG} We describe the collective spin dynamics in terms of a smoothly-varying SO(3) order parameter, subjected to topological constraints similar to those of $^3$He-A\cite{Bhattacharyya_HO_Mermin} and the $S=1$ ferromagnetic state of spinor condensates.\cite{Ho} The theory also applies to a broader class of magnetically frustrated materials.\cite{Dombre_Read,frustration1,frustration2}

\begin{figure*}[t!]
\begin{center}
\includegraphics[width=0.9\textwidth]{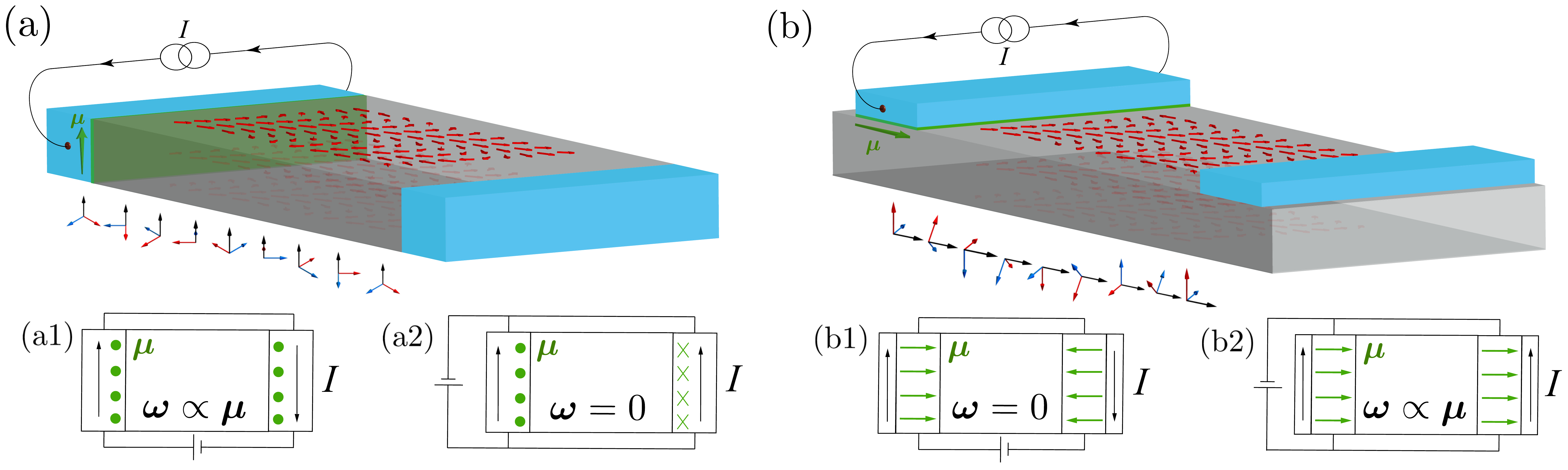}
\caption{Two-terminal geometries for the generation and detection of coherent spin transport in amorphous magnets. The precession of the CSG state (red arrows) along the spin accumulation $\bm{\mu}$ (green arrows) is depicted as a rotating triad of vectors, which represents the internal spin frame of the texture adapted to the instantaneous state of the magnet. In the case of lateral contacts, there is a nonlocal correction to the effective resistivity of the metals when the external circuit is closed in series (a1), whereas the texture remains static when the circuit is closed in parallel (a2). This is just the other way around when the contacts are on top of the sample, since the sign of the drag changes in that case.}
\vspace{-0.5cm} 
\label{Fig2}
\end{center}
\end{figure*}

\subsection{Correlated spin glass (CSG)}

Before we outline our main findings, let us describe the physical scenario that we have in mind. The microscopic interactions in amorphous magnets, particularly in rare-earth transition-metal compounds, are usually described by the Hamiltonian \cite{Harris1}\begin{align}
\label{eq:Harris}
& H=-J\sum_{\left\langle i j \right\rangle}\mathbf{\hat{S}}_i\cdot\mathbf{\hat{S}}_j-D\sum_{i}\left[\bm{\zeta}\left(\vec{r}_i\right)\cdot\mathbf{\hat{S}}_i\right]^2,
\end{align}
where $\mathbf{\hat{S}}_i$ are the spin operators at positions $\vec{r}_i$, separated by the atomic distance $a$ (the lattice constant of the original, crystallographic material). $J$ and $D$ measure the strength of the exchange and anisotropy couplings, respectively. The sign of $J$ is not relevant at macroscopic scales as long as it stabilizes a collinear state (ferro- or N\'{e}el antiferromagnetic) in the parent crystalline material below the ordering temperature, $T_c\sim J$. For our discussion, let us assume $J>0$, so the exchange interaction tends to order the spins (red arrows in Fig.~\ref{Fig1}) ferromagnetically. The unit-length vectors $\bm{\zeta}\left(\vec{r}_i\right)$ indicate the direction of the local anisotropy axis (dashed lines in Fig.~\ref{Fig1}) defined by the atomic arrangement around $\vec{r}_i$. We assume that the structural disorder of the magnet remains quenched and that $\bm{\zeta}$ follows a random distribution with no special preferred direction, $\langle\bm{\zeta}\rangle=0$. The components of these vectors can be correlated over few atomic sites, within the crystal grains of typical size $R_a$, $\langle\zeta_{\alpha}(\vec{r}_i)\zeta_{\beta}(\vec{r}_j)\rangle  \sim e^{-|\vec{r}_i-\vec{r}_j|^2/2R_a^2}\,\delta_{\alpha\beta}$.

In the crystalline material, the spins would be ordered along the uniform easy-axis, with domain walls of characteristic width $\delta_{\textrm{dw}}=\sqrt{J/D}\,a$. Nevertheless, disorder in $\bm{\zeta}$ breaks the long-range magnetic order.\cite{Imry-Ma,RG} Below the freezing temperature $T_f\lesssim T_c$, the system is generically a spin glass, characterized by a nonzero value $q$ of the Edwards-Anderson order parameter,\cite{Edwards_Anderson}
\begin{align}
\label{eq:q}
q\delta_{\alpha\beta}=\frac{1}{N}\sum_i\left\langle \hat{S}_i^{\alpha} \right\rangle \left\langle\hat{S}_i^{\beta} \right\rangle,
\end{align}
where $N$ is the total number of spins and $\langle\hat{\varsigma}\rangle$ denotes a statistical average. We can distinguish two different glassy phases arising from the competition between the two mesoscopic length scales of the model, $R_a$ and $\delta_{\textrm{dw}}$. When $R_a>\delta_{\textrm{dw}}$, the exchange interaction is predominantly frustrated and the local magnetic moments remain pinned to the local anisotropy axis, as sketched in Fig.~\ref{Fig1}(a). These textures receive the name of \textit{speromagnet}.\cite{speromagnet} In the opposite limit, $\delta_{\textrm{dw}}>R_a$, the CSG spin texture is smooth on the scale of the correlation length $R_c\sim R_a \,(\delta_{\textrm{dw}}/R_a)^{4/(4-d)}>R_a$, with $d$ being the dimensionality of the system.\cite{CSG0,CSG}  
As depicted in Fig.~\ref{Fig1}(b), a uniform, collective rotation of the CSG texture connects physically distinguishable states with approximately the same energy. On the contrary, rotations of the speromagnet always cost energy. We are thus interested in the CSG regime appearing at not too low temperatures below $T_f$, for which the magnetic medium is expected to respond elastically to external perturbations.\cite{Gardner}

\subsection{Main results}

Our main findings are synthesized in the equations of motion describing the hydrodynamic flow of spin angular momentum in the CSG phase:
\begin{subequations}
\label{eq:motion}
\begin{align}
\label{eq:gen6}
\bm{\omega}&=\chi^{-1}\,\bm{m},\\
\label{eq:gen5}
\partial_t\bm{m}-\mathcal{A}\vec{\nabla}\cdot\vec{\bm{\Omega}}+\alpha s\bm{\omega}&=\frac{g}{4\pi}\left(\bm{\mu}-\hbar\bm{\omega}\right)\delta\left(x\right).
\end{align}
\end{subequations}
Equation~\eqref{eq:gen6} provides the constitutive relation between the nonequilibrium spin density $\bm{m}$ and the angular velocity $\bm{\omega}$ of the order parameter, where the spin susceptibility $\chi$ plays the role of the moment of inertia. Equation~\eqref{eq:gen5} must be interpreted as the continuity equation for the spin density, accounting for losses due to dissipative processes in the bulk (parametrized by the Gilbert damping constant $\alpha$ in the parent crystallographic material,\cite{Gilbert} where $s\approx \hbar S/a^d$ and $S$ is the length of the microscopic spin operators) and at the interface (located at $x=0$ for concreteness). The latter is described by the source term on the right-hand side, where $\bm{\mu}$ is the spin accumulation in the metal and $g$ is a generalized spin-mixing conductance.\cite{spin-transfer} The spin current is found to be 
\begin{align}
\label{eq:spincurr}
\vec{\bm{J}}=-\mathcal{A}\,\vec{\bm{\Omega}}=-\frac{i\mathcal{A}}{2}\,\text{Tr}\left[\hat{R}^T\bm{\hat{L}}\,\vec{\nabla}\hat{R}\right],
\end{align}
where $\vec{\bm{\Omega}}$ describes the spatial variation of the collective spin rotation $\hat{R}$ defining the instantaneous state of the magnet and $\mathcal{A}\approx Ja^{2-d}$ is the stiffness of the order parameter, which maintains the spatial coherence of such rotation along the sample. Here, $\bm{\hat{L}}$ is a vector containing the generators of SO(3) with matrix elements $[\hat{L}_{\alpha}]_{\beta\gamma}=-i\epsilon_{\alpha\beta\gamma}$. 

We apply this set of equations to the device geometry usually utilized in nonlocal transport measurements.\cite{exp,vanWees} As depicted in Fig. \ref{Fig2}, we focus on two specific configurations defined by whether the heavy-metal contacts are deposited on the lateral sides of the magnet [panel (a)] or on top of it [panel (b)]. In open geometries, the precession of the spin texture manifests itself as a drag signal decaying algebraically with the length of the film, in contrast to the exponential decay of (incoherent) magnon currents.\cite{vanWees} In linear response, the heterostructure is characterized by the resistivity (defined by the ratio of the detected voltage, per unit length, to the injected current density)
\begin{equation}
\label{eq:drag}
\varrho_{\textrm{drag}}=\left(\hat{n}_i\cdot\hat{n}_d\right)\,\frac{\vartheta_{i}\vartheta_{d} R_{Q}}{t_ d\left(g_i+g_d+\frac{4\pi \alpha sL_{t}}{\hbar}\right)},
\end{equation}
where $R_Q=h/2e^2\approx 12.9$ k$\Omega$ is the quantum of resistance, $\vartheta_{i(d)}$ is the spin Hall angle in the injector (detector) metal, $t_d$ is the thickness of the detector strip, and $L_{t}$ is the distance between terminals. The prefactor determines the sign of the drag effect, which is negative for lateral terminals ($\hat{n}_{i}=-\hat{n}_{d}=\hat{x}$) and positive for terminals on top ($\hat{n}_{i}=\hat{n}_{d}=\hat{z}$). Deviations in the sign might reveal the presence of parasitic signals, like current leakage from the injector to the detector.

In close geometries (insets of Fig. \ref{Fig2}), the coherent spin dynamics induces a nonlocal correction to the effective resistivity that depends on the configuration of the external circuit. For Pt contacts, the resultant magnetoresistance is about 10\% of their resistivity at room temperature. Furthermore, we argue that in short enough devices the spin currents are stable due to the topology of the order-parameter manifold, the group of proper rotations; in particular, current states are classified in two distinct topological sectors. Spin supercurrents, however, are only stabilized in the thermodynamic limit by additional easy-plane anisotropies, i.e., when the random anisotropy axes $\bm{\zeta}$ lie predominantly within the plane of the film. This anisotropy may arise due to applied/growth-induced strain or electrical gating. In that case, the supercurrents decay through thermally activated $4\pi$-phase slips, which would be manifested in nonlinear interference effects.\cite{interference}

The rest of the article is structured as follows. In Sec. \ref{Macroscopic_dynamics}, we derive the Lagrangian describing the macroscopic spin dynamics in the CSG phase, from which Eqs. \eqref{eq:motion} are obtained. In Sec. \ref{Nonlocal_transport}, we apply these equations to study the linear response of the systems in the two-terminal geometry of Fig.~\ref{Fig2}. Section \ref{Topological_stability} deals with the topological stability of the spin supercurrents and their degradation by phase slips. Section \ref{Discussion} contains final conclusions and suggestions for experiments. Some technical details are given in Appendices \ref{AppA} and \ref{AppB}.

\section{Macroscopic dynamics}
\label{Macroscopic_dynamics}

The dynamics in the CSG phase is glassy, characterized by a rough landscape of free-energy minima.\cite{CSG2} We are interested only in nonequilibrium macroscopic deviations for which the system remains within a given (local) minimum basin. The latter is defined by the initial state $G$ of (mutual) equilibrium of the magnet in contact with metallic reservoirs and negligible macroscopic magnetization. From this point forward, we approximate the statistical averages in Eq.~\eqref{eq:q} by the (quantum-mechanical) expectation value $\langle\, \hat{\varsigma}\,\rangle_G=\text{Tr}(\hat{\varsigma}\,\hat{\rho}_G)$, where $\hat{\rho}_G$ represents the density-matrix operator of state $G$.

Following the conventional program in hydrodynamics,\cite{Chaikin_Lubensky} we also consider states $G'=gG$ generated by the symmetry operations that connect physically distinguishable spin configurations with the same energy,\cite{Mermin} i.e. the group of proper rotations in the present case, $g\in$ SO(3). Note that there may be other states $G''$ with approximately the same free energy that are not connected to $G$ by a proper rotation, for example, a spatial or time inversion. We assume that these states are disconnected by large free-energy barriers, so if the system is initiated in $G$, there is a negligible probability of reaching a different minimum basin $G''$. Macroscopic deviations from equilibrium are described then by $\hat{\rho}_{\textrm{neq}}(t,\vec{r})=\hat{\mathcal{U}}(t,\vec{r})\hat{\rho}_{G}\,\,\hat{\mathcal{U}}^{\dagger}(t,\vec{r})$, where $\hat{\mathcal{U}}(t,\vec{r})$ is a slowly-varying (in the scale of $R_c$) SU(2) spin rotation.

\subsection{Coarse-grained Lagrangian}

Following Halperin and Saslow,\cite{Halperin_Saslow} we introduce the one-body operator
\begin{align}
\hat{\mathcal{R}}_{\alpha\beta}\left(\vec{r}\right)\equiv\frac{1}{q N_{\vec{r}}}\sum_{i\in \mathcal{V}_{\vec{r}}}\left\langle \hat{S}_i^{\beta}\right\rangle_G \hat{S}_i^{\alpha}.
\end{align}
Here, $N_{\vec{r}}$ is the number of spins contained in $\mathcal{V}_{\vec{r}}\gtrsim (R_c)^d$, the volume element around $\vec{r}$ defining our coarse-graining procedure. Note that we have divided the above operator by the equilibrium value of the Edwards-Anderson order parameter introduced in Eq. \eqref{eq:q}. For smooth deviations we have $\langle\hat{\mathcal{R}}_{\alpha\beta}(\vec{r})\rangle_{\textrm{neq}}= R_{\alpha\beta}(t,\vec{r})+\mathcal{O}\big(1/\sqrt{N_{\vec{r}}}\hspace{0.05cm}\big)$, where $R_{\alpha\beta}(t,\vec{r})$ are the matrix elements of the SO(3) rotation associated with $\hat{\mathcal{U}}(t,\vec{r})$. In order to describe the dynamics of this order parameter, we have to introduce also auxiliary fields related to the infinitesimal generators of spin rotations,
\begin{align}
\bm{\hat m}\left(\vec{r}\right)\equiv\frac{\hbar}{\mathcal{V}_{\vec{r}}}\sum_{i\in\mathcal{V}_{\vec{r}}}\mathbf{\hat{S}}_i.
\end{align}
By construction, the macroscopic spin density is zero at equilibrium, $\left\langle\bm{\hat{m}}\left(\vec{r}\right)\right\rangle_{G}=0+\mathcal{O}\big(1/\sqrt{N_{\vec{r}}}\hspace{0.05cm}\big)$, while nonequilibrium deviations, $\langle\bm{\hat{m}}(\vec{r})\rangle_{\textrm{neq}}\equiv \bm{m}\left(t,\vec{r}\right)$, vary smoothly on the scale of $\mathcal{V}_{\vec{r}}$, $\left|\bm{m}\left(\vec{r}\right)\right|\ll \hbar S/\mathcal{V}_{\vec{r}}$.

The dynamics of these variables is governed by the phase-space Lagrangian density
\begin{align}
\label{eq:Lagrangian}
\mathcal{L}\big[\bm{m},\hat{R}\big]=\bm{m}\cdot\bm{\omega}-\frac{\mathcal{A}}{4}\,\text{Tr}\left[\partial_\mu\hat{R}^T\partial_\mu\hat{R}\right]-\frac{\left|\bm{m}\right|^2}{2\chi},
\end{align}
where summation over repeated spatial indices ($\mu$) is implicit. The first term enforces the conjugacy relations between $\hat{R}$ and $\bm{m}$, the latter playing the role of the canonical angular-momentum density, $\bm{m}\equiv\partial\mathcal{L}/\partial\bm{\omega}$.\cite{Marchenko} The angular velocity is defined through the equation of motion for the unitary rotation, $i\partial_t\hat{\mathcal{U}}=(\bm{\omega}\cdot\mathbf{\hat{S}})\,\hat{\mathcal{U}}$; introducing the associated SO(3) matrices and inverting this relation yields the compact expression\begin{align}
\bm{\omega}=\frac{i}{2}\,\text{Tr}\left[\hat{R}^T\bm{\hat{L}}\,\partial_t\hat{R}\right],
\end{align}
Note that the conjugacy relations between our hydrodynamical variables derived from the above Lagrangian correspond to the classical limit $\{\varsigma_{1}(\vec{r}),\varsigma_{2}(\vec{r}\,')\}\equiv-i/\hbar\,\langle[\hat{\varsigma}_{1}(\vec{r}),\hat{\varsigma}_{2}(\vec{r}\,')] \rangle_{\textrm{neq}}$ of the commutation relations between local (coarse-grained) quantum operators: 
\begin{align}
\label{eq:Poisson}
 \left\{m_{\alpha}\left(\vec{r}\right),R_{\beta\gamma}\left(\vec{r}\,'\right)\right\}&\approx \epsilon_{\alpha\beta\lambda}\,R_{\lambda\gamma}\left(\vec{r}\right)\,\delta\left(\vec{r}-\vec{r}\,'\right).
\end{align}

The last two terms in Eq.~\eqref{eq:Lagrangian} corresponds to a phenomenological expansion (up to quadratic order) of the free-energy cost of deviations from equilibrium within a given minimum basin,\cite{foot} which, for the CSG, can be coarse-grained from the Hamiltonian in Eq.~\eqref{eq:Harris}, adhering to the hierarchy of length scales sketched in Fig.~\ref{Fig1}. Spatial variations of $\hat{R}(t,\vec{r})$ has a cost in exchange energy provided by the stiffness of the SO(3) order parameter, $\mathcal{A}$, ultimately related to $T_f$ (and therefore corresponding to a fraction of $Ja^{2-d}$). The last term accounts for the free-energy cost of a macroscopic (on a scale larger than $R_c$) spin density in the CSG phase, inversely proportional to its spin susceptibility,\cite{CSG}\begin{align}
\chi\approx \frac{\hbar^2(R_c/R_a)^{d/2}}{Da^d}.
\end{align}
Note that the last term incorporates the effect of both the strength $D$ and spatial distribution (through $R_a$) of the random anisotropy.


Integration out of the slave variable $\bm{m}$ in Eq.~\eqref{eq:Lagrangian} yields the following Lagrangian of a O(4) nonlinear $\sigma$-model,
\begin{align}
\label{nonlinsigmamodel}
L=\frac{1}{4}\int\,d\vec{r}\left(\chi\,\text{Tr}\left[\partial_t\hat{R}^T\partial_t\hat{R}\right]-\mathcal{A}\,\text{Tr}\left[\partial_\mu \hat{R}^T\,\partial_\mu\hat{R}\right]\right),
\end{align}
which also describes the macroscopic dynamics of multi-lattice antiferromagnets in frustrated lattices.\cite{Dombre_Read,frustration1,frustration2} In its linearized version, this model yields three independent soft modes with velocity $c=\sqrt{\mathcal{A}/\chi}$. \cite{Halperin_Saslow,Marchenko}

\subsection{Dissipation and interfacial torques}

Dissipation can be introduced by means of a Rayleigh function $\mathcal{R}$. We cast the power density dissipated in the bulk of the magnet as a quadratic form in $\bm{\omega}$, $P_{\textrm{bulk}}=2\,\mathcal{R}_{\textrm{bulk}}=\alpha s\bm{\omega}^2$. The dimensionless parameter $\alpha$ can be assumed to be close to the Gilbert damping constant\cite{Gilbert} in the parent crystallographic material. We also consider spin-transfer torques and enhanced dissipation at the interface with a normal metal. The interfacial dissipation rate per unit of area can be generically written as $\bar{P}_{\textrm{int}}=\bm{\omega}^T\hat{g}\,\bm{\omega}$, where $\hat{g}$ is a symmetric 3$\times$3 matrix parametrizing the heat flow from the magnet into the metal.\cite{spin-transfer} Diagonalization of this matrix provides three non-negative damping parameters associated with the rotations along the principal axes of the interface, which define a natural laboratory frame to study the spin dynamics. The eigenvalues of the kernel $\hat{g}$ generalize the concept of spin-mixing conductance and, like in the case of collinear magnets, they admit a microscopic expression in terms of the reflection coefficients of the interface.\cite{spin-transfer} 

In the presence of a nonequilibrium spin accumulation $\bm{\mu}$, the energy flow across the interface is modified by the work exerted by itinerant electrons on the magnetic system: we have to substitute $\bm{\omega}$ by $\bm{\omega}-\bm{\mu}/\hbar$ in the expression for $\bar{P}_{\textrm{int}}$, since the system is in a state of mutual dynamic equilibrium when $\hbar\bm{\omega}=\bm{\mu}$.\cite{Tserkovnyak_Brataas} In the limit of exchange-dominated interactions we assume isotropy in spin space (as in Eq.~\ref{eq:Lagrangian}), $\hat{g}= g\hat{1}$, and hence the interfacial Rayleigh function for the CSG phase becomes
\begin{align}
\bar{\mathcal{R}}_{\textrm{int}}=\frac{\hbar g}{8\pi}\left(\bm{\omega}-\frac{\bm{\mu}}{\hbar}\right)^2.
\end{align}


\section{Nonlocal transport}
\label{Nonlocal_transport}

The equations of motion \eqref{eq:motion} are derived from the modified variational principle $\delta_{\varsigma}\mathcal{L}=\delta_{\dot{\varsigma}}\mathcal{R}$. Integrating Eq.~\eqref{eq:gen5} over an infinitesimal volume around the interface generates the boundary condition for the spin current,
\begin{align}
\hat{n}\cdot\vec{\bm{J}}=\frac{g}{4\pi}\left(\bm{\mu}-\hbar\bm{\omega}\right),
\end{align}
where $\hat{n}$ denotes the normal vector (to the metallic interface) inwards the magnet.

We apply now these equations to the device geometry of Fig.~\ref{Fig2}. Note first that we may interpret $\bm{m}/\chi$ in the right-hand side of Eq.~\eqref{eq:gen6} as the analog of the chemical potential in the Josephson frequency relation of mass superfluids. Therefore, the angular velocity must be uniform and constant in the steady state. In the open configuration, the charge current $\vec{j}$ flowing within the left terminal (injector) induces a nonequilibrium spin accumulation $\bm{\mu}$ at the interface via the spin Hall effect, setting a coherent precession of the disordered texture. The spin accumulation must be determined self-consistently by solving the charge/spin continuity equations at the metal subjected to suitable boundary conditions. We expect the spin accumulation to be exponentially localized at the interface, in a length scale of the order of the spin-diffusion length in the metal that we assume much shorter than the terminal thickness. In this limit, and assuming that the metal behaves as a perfect spin sink, the spin current injected into the magnet reads\cite{spinHall} \begin{align}
\hat{n}_i\cdot\vec{J}_{\alpha}=\frac{\hbar \vartheta_{i}}{2e}\left(\hat{n}_i\times\vec{j}\right)_{\alpha}-\frac{\hbar g}{4\pi}\omega_{\alpha}.
\end{align}
With no external bias applied to the right terminal (detector), Onsager reciprocity dictates the onset of a (charge-pumping) electromotive force of the form $\mathcal{E}_i=(\hbar\,\vartheta_d)/2et_d)(\bm{\omega}\times\hat{n}_{d})_i$.\cite{spinHall} The heterostructure is then characterized by the drag resistivity in Eq.~\eqref{eq:drag}. By taking typical values of $\alpha=10^{-4}$, $s/\hbar=10^{28}$ m$^{-3}$, $\vartheta=0.1$ and $g_{L/R}=10^{18}$ m$^{-2}$ for Pt$\vert$YIG interfaces,\cite{interface_numbers} we estimate $|\varrho_d|\approx10^{-2}$ $\mu\Omega\cdot$cm for $t\approx10$ nm and $L_{t}=10$ $\mu$m. As a result, nonlocal voltage signals in the range of $V_{\textrm{nl}}\approx 0.1$ mV could be achieved for Pt-contact lengths of $1$ mm and $j=10^9$ A/m$^{2}$, the current densities applied in Ref.~\onlinecite{exp}.

Alternatively, the external circuit can be closed, as sketched in the insets of Fig.~\ref{Fig2}, leading to a nonlocal magnetoresistance. When the spin accumulations are opposite, the texture remains static and there is no correction to the effective resistivity of the metals. On the contrary, if the spin accumulations are parallel, the pumping electromotive force in favor of the external battery reduces the effective resistivity by $\rho_{m}=-2|\rho_{\textrm{drag}}|=-\vartheta^{2}R_{Q}/t(g+2\pi\alpha sL_{t}/\hbar)$ (assuming identical interfaces for simplicity). From the previous estimates we obtain $\rho_{m}\sim\mu\Omega\cdot$cm for $L_{t}\ll\hbar g/2\pi\alpha s\simeq0.1\mu$m.

\section{Topological stability}
\label{Topological_stability}

The onset of a coherent spin precession does not exclude the possibility of spin-current degradation by (thermal) fluctuations. We can speak of spin supercurrents only if there is an energy gap of topological origin that precludes the relaxation of the current into a uniform state. In this section, we analyze the topology of the order-parameter manifold SO(3) and the proliferation of phase slips in macroscopic devices.

\subsection{Order-parameter manifold}

\begin{figure}[t!]
\begin{center}
\includegraphics[width=1\columnwidth]{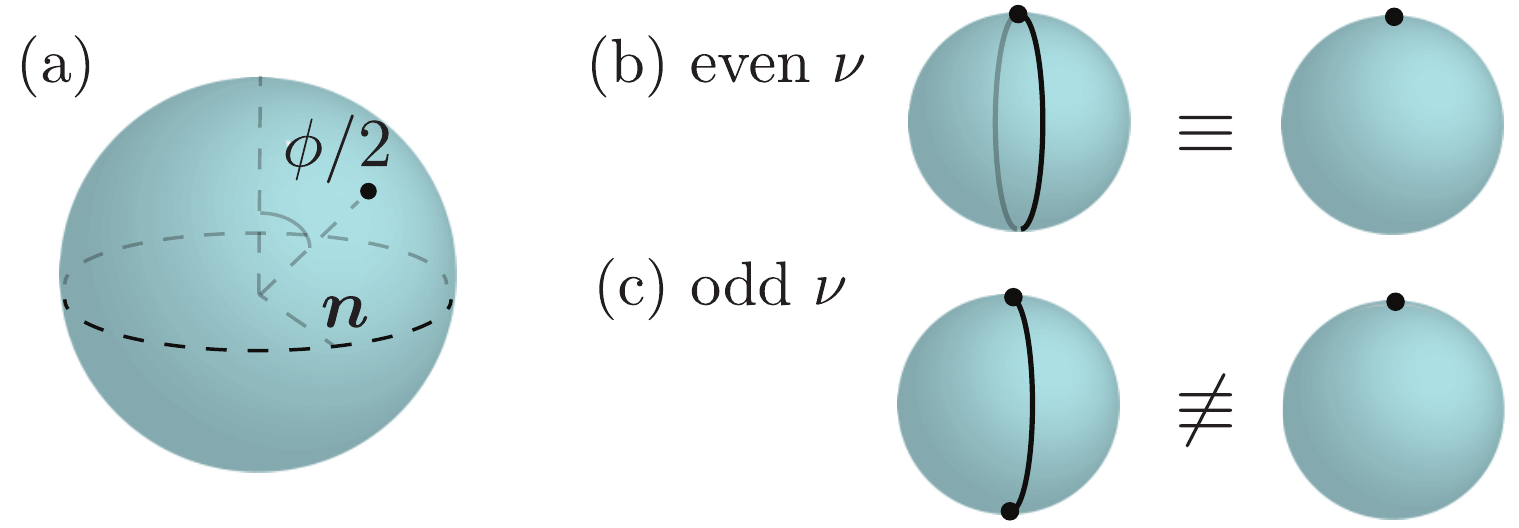}
\caption{a) Unit-radius cross sections of the quaternion hypersphere represent all possible rotations by angle $\phi$ around $\bm{n}$. Antipodal points correspond to equivalent rotations, $\left(\phi,\bm{n}\right)$ and $\left(2\pi-\phi,-\bm{n}\right)$. b) Current states of even winding are mapped to loops starting and ending at the same point. The texture can be smoothly deformed into the ground state. c) Current states of odd winding are mapped to loops starting and ending at antipodal points. The minimum winding $|\nu|=1$ cannot relax (smoothly) into the ground state.}
\vspace{-0.5cm} 
\label{Fig3}
\end{center}
\end{figure}

The topology of SO(3) is better understood through quaternions: a proper rotation $\hat{R}$ is represented by two 4-dimensional vectors $\mathbf{q}=(w,\bm{v})$ and $-\mathbf{q}$ satisfying $\omega^2+|\bm{v}|^2=1$, see Appendix \ref{AppA}. The SO(3) manifold is then homeomorphic to the topological space of lines passing through the origin in $\mathbb{R}^4$, or equivalently, the unit hypersphere with antipodal points being identified as the same. This hypersphere can be depicted via spherical sections, where one of the components remains constant. Quaternions lying on this section are of the form $\mathbf{q}=(\cos\phi/2,\sin\phi/2\,\bm{n})$, describing physical spin rotations by angle $\phi$ around axis $\bm{n}$, see Fig.~\ref{Fig3}(a).

\begin{figure*}
\begin{center}
\includegraphics[width=0.9\textwidth]{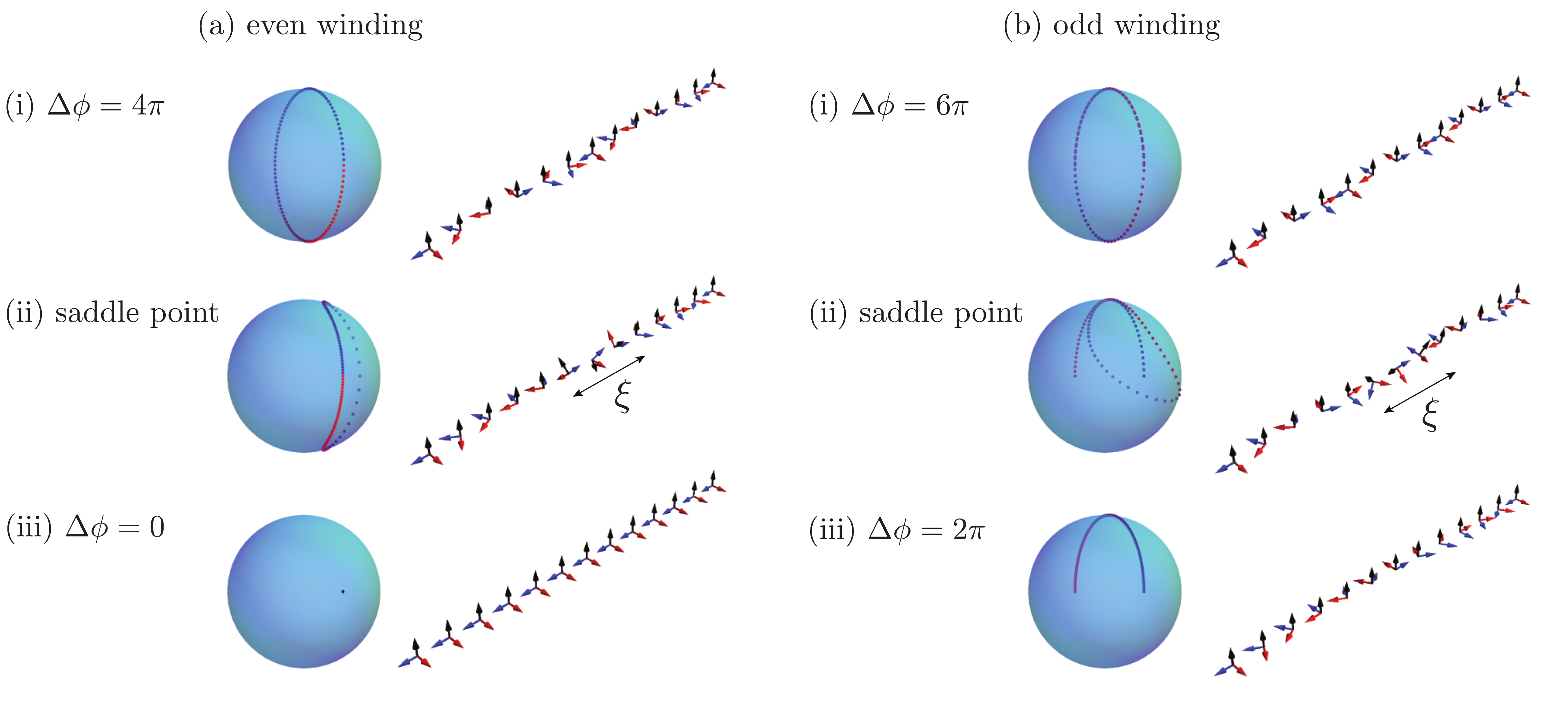}
\caption{Degradation of spin supercurrents in amorphous magnets. In panel (a), a supercurrent state with even winding (i) decays to the ground state (iii) through a $4\pi$-phase slip (ii) taking place in the macroscopic length scale defined by the remanent easy-plane anisotropy. Panel (b) shows the analogous process for supercurrents with odd winding, which only relax into the ground state by proliferation (nucleation and expansion) of disclinations in the order parameter ($\mathbb{Z}_2$ vortices shown in Fig.~\ref{Fig5}).}
\vspace{-0.5cm} 
\label{Fig4}
\end{center}
\end{figure*}

The coherent precession of the spin texture along a fixed axis (defined by the spin accumulation $\bm{\mu}$ in the adjacent metals) can be mapped to a geodesic loop, as represented in Fig.~\ref{Fig3}(b)~and~(c). We consider fixed boundary conditions in the geometry of Fig.~\ref{Fig2}(a). When the external circuit is closed in parallel, the spin texture at the left/right terminal remains fixed to a (quasi-)static state of mutual equilibrium with the lateral contacts. The internal spin frame of the texture (represented by a triad in Fig.~\ref{Fig2}) rotates by an angle $\Delta\phi=2\pi\nu$ between terminals, where $\nu$ is the winding of the corresponding rotation. States with even winding number (i.e., rotating a multiple of $4\pi$) correspond to loops beginning and ending at the same point, as illustrated in Fig.~\ref{Fig3}(b). These loops can be smoothly deformed (i.e., there are topologically equivalent) to a single point, and therefore, they always relax into the ground state in the absence of additional anisotropies. Current states with odd winding correspond to loops beginning and finishing at antipodal points, as represented in Fig.~\ref{Fig3}(c). This constraint implies that a state of winding $\nu=2n+1$ can decay to a state with winding $\nu=2n-1$, with $n$ an integer, but not to the ground state. This parity distinction is traced to the fundamental group of the order parameter, $\pi_1\left(\text{SO(3)}\right)=\mathbb{Z}_2$.

Since the order-parameter manifold is not simply connected, we could argue that in short enough devices a spin current of the order of $2\pi\,\mathcal{A}/L_{t}$ ($|\nu|=1$) is stable. Supercurrent states in the thermodynamic sense, however, are only stable in the presence of additional easy-plane anisotropies, as we analyze next.

\subsection{Phase slips}

In thin films like the ones considered in Fig.~\ref{Fig2}, rotations that remove the spins from the plane of the film may have an extra free-energy cost,
\begin{align}
\mathcal{F}_{\textrm{an}}=\frac{\mathcal{K}}{2}\left(v_x^2+v_y^2\right)=\frac{\mathcal{K}}{2}\sin^2\frac{\beta}{2}.
\end{align}
The last expression corresponds to a parametrization of SO(3) matrices in terms of proper Euler angles,
\begin{align}
\hat{R}[\alpha,\beta,\gamma]=e^{-i\alpha \hat{L}_z}e^{-i\beta\hat{L}_y}e^{-i\gamma \hat{L}_z}.
\end{align}
The total (static) free-energy density of the CSG reads then
\begin{align}
\label{eq:free-energy}
 \mathcal{F}\big[\phi,\theta,\chi\big]= &  \frac{\mathcal{A}}{2}\left[\sin^2\theta\big(\vec{\nabla}\phi\big)^2+\cos^2\theta\big(\vec{\nabla}\chi\big)^2\right]\\
& +2\mathcal{A}\big(\vec{\nabla}\theta\big)^2+\frac{\mathcal{K}}{2}\,\cos^2\theta,\nonumber
\end{align}
where we have introduced the following angular variables: $\phi\equiv\alpha+\gamma$, $\chi\equiv\gamma-\alpha$, and $\theta\equiv(\pi-\beta)/2$. Only spin rotations of the form $\hat{R}_z[\phi]=e^{-i\phi\hat{L}_z}$ are soft, whereas the other two modes develop a gap, $\omega_{x,y}=c\,\sqrt{|\vec{q}|^2+\xi^{-2}}$; here $\xi=2\sqrt{\mathcal{A}/\mathcal{K}}$ is a characteristic length scale associated with the remanent anisotropy.

Consider, for example, the situation in Fig.~\ref{Fig2}(a). The $z$-spin supercurrent injected by the spin accumulation $\bm{\mu}\propto\bm{z}$, $\vec{J}_z=-\mathcal{A}\vec{\nabla}\phi$, becomes energetically unstable when the superfluid phase changes in space faster than $2/\xi$. This criterion defines the Landau critical current\begin{align}
\label{eq:Landau}
\big|\vec{J}_z^c\big|=2\mathcal{A}/\xi=\sqrt{\mathcal{A}\mathcal{K}},
\end{align}
above which the energy barrier for the proliferation of smooth (i.e., coreless) phase slips goes to zero, see Appendix~\ref{AppB}. Figure~\ref{Fig4} shows the most probable phase-slip events in the case of long terminal separation, $L_t\gg \xi$, consisting of excursions of the order parameter along a spherical section parametrized by a constant value of $\chi$. The changes in phase take place on a scale of $\xi$ and are always a multiple of $4\pi$.

\section{Discussion}
\label{Discussion}

\begin{figure}
\begin{center}
\includegraphics[width=\columnwidth]{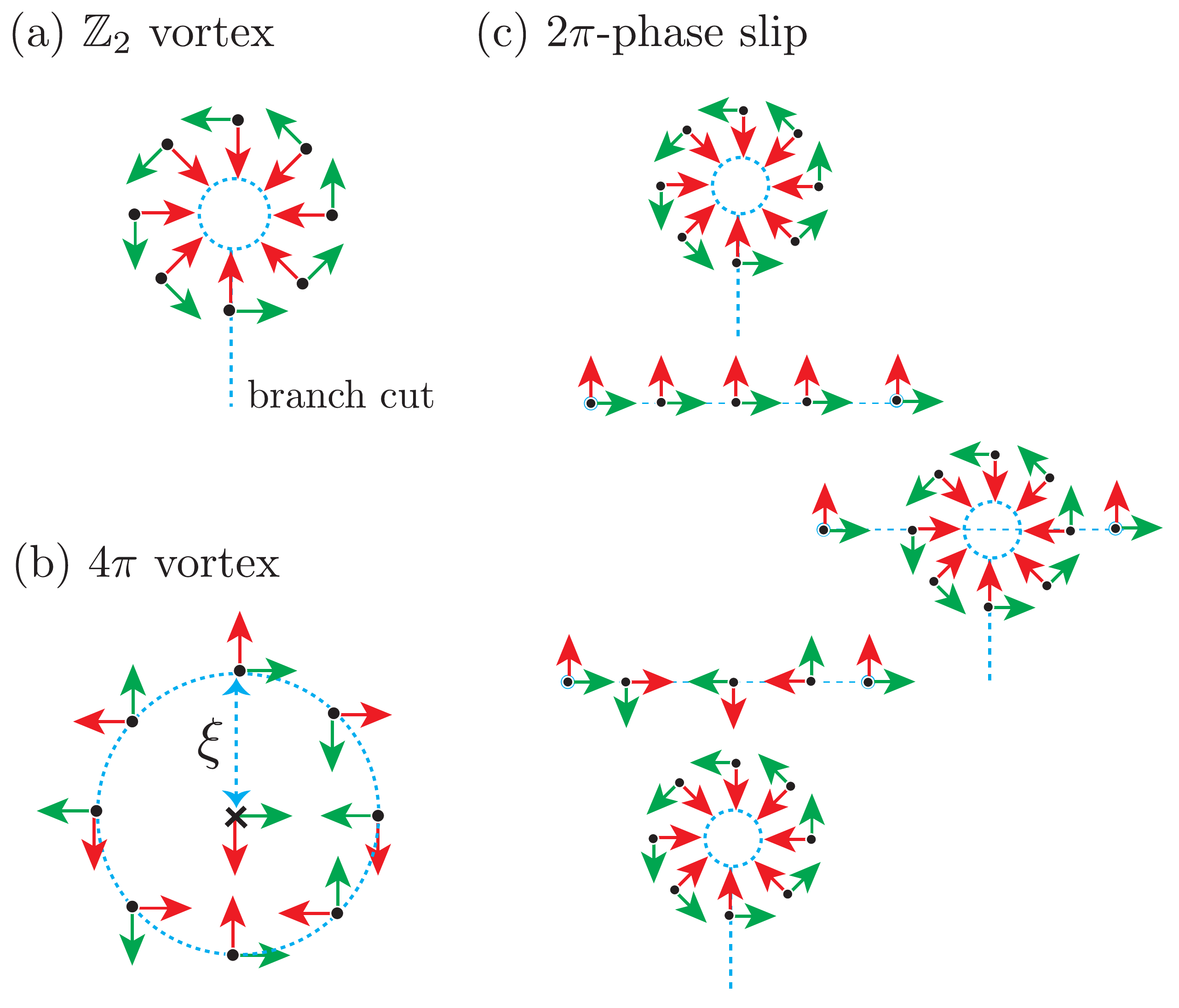}
\caption{(a) $\mathbb{Z}_2$ (singular) vortex. The dashed lines represents the essential branch cut where the SO(3) order parameter (represented as a tetrad of vectors) is multivalued (antipodal points of the SO(3) hypersphere are identified there). (b) $4\pi$ (\textit{coreless}) vortex. The singularity is avoided by smooth rotation of the texture on the scale of $\xi$. (c) A $\mathbb{Z}_2$ vortex crossing the superfluid streamlines induces $2\pi$-phase slips.}
\vspace{-0.5cm} 
\label{Fig5}
\end{center}
\end{figure}

Landau's criterion is a necessary but not sufficient condition for the stability of the spin superfluid. Phase slips can be thermally activated and monitored as jumps in the magnetoresistance when the external circuit is closed in parallel (in series for the {\it on-top} configuration).\cite{slips_SeKwon} On the other hand, jumps of $2\pi$ in the superfluid phase are only possible due to the proliferation (nucleation and expansion) of vortex disclinations in the order parameter, as represented in Fig.~\ref{Fig5}. These topological defects are characterized by a $\mathbb{Z}_2$ charge, expressing a fundamental difference between odd and even vorticity: in the former case, the SO(3) order parameter is not properly defined within the core, while in the latter case the singularity in $\phi$ is avoided by a smooth rotation of the texture, like the $4\pi$-vortices in $^{3}$He-A.\cite{Anderson_Tolouse} The core radius corresponds to a mesoscopic scale not captured by the macroscopic Lagrangian in Eq.~\eqref{eq:Lagrangian}. When the superfluid phase changes on lengths comparable to this mesoscopic scale, the system is no longer robust against the proliferation of disclinations, or in other words, our coarse-graining procedure breaks down.

In conclusion, we have stablished the nonlinear equations governing the macroscopic spin dynamics of insulating amorphous magnets in contact with metallic reservoirs. The onset of a coherent precession of a smoothly disordered, noncollinear spin texture can be detected as a long-ranged drag signal and related magnetoresistance effects in nonlocal transport. A remanent easy-plane anisotropy in the CSG state stabilizes spin supercurrents in the thermodynamic limit. These currents decay through thermally activated $4\pi$-phase slips, characteristic of the emergent SO(3) order parameter. The $4\pi$- vs. conventional $2\pi$-phase slips can be revealed through the superfluid interference in a loop geometry, doubling the periodicity of the critical current as a function of the control parameters, as discussed in Ref.~\onlinecite{interference}.

\begin{acknowledgments}

We would like to thank Pramey Upadhyaya and Eugene Chudnovsky for valuable discussions. This work has been supported by U.S. Department of Energy, Office of Basic Energy Sciences under Award No. DE-SC0012190. R.Z. thanks Fundaci\'{o}n Ram\'{o}n Areces for support through a postdoctoral fellowship within the XXVII Convocatoria de Becas para Ampliaci\'{o}n de Estudios en el Extranjero en Ciencias de la Vida y de la Materia.

\end{acknowledgments}

\appendix

\section{Quaternion representation}
\label{AppA}

Unit-norm quaternions (so-called versors) provide a convenient parametrization of rotation matrices: since SU(2) is the universal (double) covering of SO(3) and also isomorphic to the unit hypersphere in $\mathbb{R}^{4}$, we can represent a generic SO(3) rotation via a 4-component unit vector $\mathbf{q}=(w,\bm{v})$ according to
\begin{align}
\label{eq:def_quaternion}
\hat{\mathcal{U}}=w\hat{1}-i \bm{v}\cdot\bm{\sigma}:= w\hat{1}-ix\hat{\sigma}_x-iy\hat{\sigma}_y-iz\hat{\sigma}_z,
\end{align}
where $\bm{\sigma}=(\hat{\sigma}_{x},\hat{\sigma}_{y},\hat{\sigma}_{z})$ is the vector of Pauli matrices and $\bm{v}=(x,y,z)$ denotes the vectorial (imaginary) part of the quaternion. Note that $\det\,\hat{\mathcal{U}}=1$ is equivalent to $w^2+\bm{v}^2=1$.  The SO(3) matrix $\hat{R}$ associated with $\hat{\mathcal{U}}\in\,$SU(2) reads as 
\begin{align}
\label{eq:RU}
R_{\alpha\beta}&=\frac{1}{2}\text{Tr}\left[\hat{\sigma}_{\alpha}\,\hat{\mathcal{U}}\hat{\sigma}_{\beta}\,\hat{\mathcal{U}}^{\dagger}\right]\\
\label{eq:quaternion_parametrization}
&=\left(1-2\left|\bm{v}\right|^2\right)\delta_{\alpha\beta}+2\,v_{\alpha}v_{\beta}-2\,\varepsilon_{\alpha\beta\gamma}\,w\,v_{\gamma}.
\end{align}
Note that both $\mathbf{q}$ and $-\mathbf{q}$ parametrize the same $\hat{R}$, so that SO(3)$\cong \mathbb{R}\textrm{P}^{3}$, i.e. $S^{3}$ with antipodal points being identified as the same. 

The set $\{\hat{1},-i\hat{\sigma}_x,-i\hat{\sigma}_y,-i\hat{\sigma}_z\}$ defines the basis of the (vector) space of quaternions $\mathbf{q}$ over the real numbers, with the usual Hamilton product:
\begin{align}
\label{eq:Hamilton}
\mathbf{q}_1\wedge\mathbf{q}_2:=\left(w_1w_2-\bm{v}_1\cdot\bm{v}_2,w_1\bm{v}_2+w_2\bm{v}_1+\bm{v}_1\times\bm{v}_2\right),
\end{align}
which is inferred directly from the algebra of Pauli matrices. Addition and multiplication by real numbers are as in $\mathbb{R}^4$. The adjoint of $\mathbf{q}=(w,\bm{v})$ is $\mathbf{q}^*=(w,-\bm{v})$, so that the norm $\sqrt{\mathbf{q}^*\wedge\mathbf{q}}$ ($=1$ in this case) is a real number. The Hamilton product provides a representation of the matrix product in SO(3): $\mathbf{q}^*$ corresponds to $\hat{R}^T$ and $\hat{R}_1\cdot\hat{R}_2$ corresponds to $\mathbf{q}_1\wedge\mathbf{q}_2$. Furthermore, the rotation of a vector $\bm{u}\in\mathbb{R}^3$ also admits a simple expression in terms of versors, $\hat{R}\cdot\bm{u}=\mathbf{q}\wedge\mathbf{u}\wedge\mathbf{q}^*$, with $\mathbf{u}=(0,\bm{u})$ denoting the embedding into the (vector) space of imaginary quaternions.

The Lagrangian \eqref{nonlinsigmamodel} can be written in terms of versors as 
\begin{align}
\label{eq2}
\mathcal{L}=2\int\,d\vec{r}\left(\chi\,\partial_t\mathbf{q}^*\wedge\partial_t\mathbf{q}-\mathcal{A}\,\partial_\mu\mathbf{q}^*\wedge\partial_\mu\mathbf{q}\right).
\end{align}
The analogy with bipartite antiferromagnets is clear by noting that the versor $\mathbf{q}$ plays the same role as the staggered magnetization $\bm{n}$ in the expressions for the spin current,
\begin{align}
\label{eq3}
\vec{\bm{J}}=2\,\mathcal{A}\,\mathbf{q}^*\wedge\vec{\nabla}\mathbf{q},
\end{align}
where the Hamilton product replaces the cross product in $\vec{\bm{J}}\sim\vec{\nabla}\bm{n}\times\bm{n}$. Analogously, the angular velocity reads as
\begin{align}
\label{eq4}
\bm{\omega}=2\,\partial_t\mathbf{q}\wedge\mathbf{q}^*.
\end{align}

\section{Local minima and saddle-point solutions}
\label{AppB}

Local minima solutions of the free energy in Eq.~\eqref{eq:free-energy} are of the form\begin{align}
\phi(s)=\phi_0+k_{\nu}s,
\end{align}
with the boundary conditions defined by the winding $\nu$ as\begin{align}
\Delta\phi=k_{\nu}\ell=2\pi\nu.
\end{align}
Here $s:= x/\xi$ is the position in between terminals and $\ell:=L_{t}/\xi$ denotes the separation in reduced units. These solutions correspond to the metastable superfluid states with persistent z-spin current $|\vec{J}_z|=\mathcal{A}k_{\nu}/\xi$.

The saddle-point solutions depicted in Fig.~\ref{Fig4} correspond to phase slips localized at the middle of a long magnetic wire. Their expressions are given by
\begin{align}
& \phi\left(s\right)=\phi_0+\bar{k}_{\nu}s+2\arctan\frac{\sqrt{4-\bar{k}_{\nu}^2}\,\tanh\frac{\sqrt{4-\bar{k}_{\nu}^2}\,s}{2}}{\bar{k}_{\nu}},\\
& \theta\left(s\right)=\arccos\left[\sqrt{1-\frac{\bar{k}_{\nu}^2}{4}}\,\text{sech}\left(\sqrt{1-\frac{\bar{k}_{\nu}^2}{4}}\,s\right)\right],\\
& \chi\left(s\right)=\chi_0,
\end{align}
where $\chi_0$ labels the spherical section and $\bar{k}_{\nu}$ satisfies the following equation inferred from the boundary conditions:
\begin{align}
\bar{k}_{\nu}\ell+4\arctan\frac{\sqrt{4-\bar{k}_{\nu}^2}\,\tanh\frac{\sqrt{4-\bar{k}_{\nu}^2}\,\ell}{4}}{\bar{k}_{\nu}}=2\pi\nu.
\end{align}
The solutions of this equation verify $k_{\nu-2}<\bar{k}_{\nu}<k_{\nu}$. Note also that the z-component of the spin current in this state reads $|\vec{J}_z|=\mathcal{A}\sin^2\theta\partial_x\phi=\mathcal{A} \bar{k}_{\nu}/\xi$. We see then that phase slips decrease the current, connecting superfluid states with winding numbers $\nu$ and $\nu-2$. The energy barriers that prevent these events can be estimated from the difference in free energy of these solutions,
\begin{align}
\Delta E_{\nu}=\frac{\mathcal{A}\hspace{0.03cm}\mathcal{S}}{\xi}\left(\bar{k}_{\nu}^2\ell-k_{\nu}^2\ell+8\sqrt{4-\bar{k}_{\nu}^2}\,\tanh\frac{\ell\sqrt{4-\bar{k}_{\nu}^2}}{4}\right),
\end{align}
where $\mathcal{S}$ is the interface area. There are other events localized around different points of the wire, but we can neglect them in our energetic analysis. These barriers vanish when
\begin{align}
\left|k_{\nu}^c\right|=\left|\bar{k}_{\nu}^c\right|=2\hspace{0.1cm}\Leftrightarrow\hspace{0.1cm}\big|\vec{J}_z^c\big|=\frac{2\mathcal{A}}{\xi}=\sqrt{\mathcal{A}\mathcal{K}},
\end{align}
which coincides with Landau's criterion in Eq.~\eqref{eq:Landau}.


\begin{thebibliography}{99}
\bibitem{foot} The Euler-Lagrange equations derived from Eq.~\eqref{eq:Lagrangian} (i.e., Eqs.~\eqref{eq:motion} with $\alpha,g=0$) can be cast as the Hamilton-Liouville set $\partial_t\bm{m}=\{\bm{m},F\}$, $\partial_t\hat{R}=\{\hat{R},F\}$, where $F$ represents the free-energy cost of smooth deviations from equilibrium, and the Poisson brackets correspond to Eq.~\eqref{eq:Poisson} and $\left\{m_{\alpha}\left(\vec{r}\right),m_{\beta}\left(\vec{r}\,'\right)\right\}\approx \epsilon_{\alpha\beta\gamma}\,m_{\gamma}\left(\vec{r}\right)\,\delta\left(\vec{r}-\vec{r}\,'\right)$, while $ \left\{R_{\alpha\beta}\left(\vec{r}\right),R_{\gamma\delta}\left(\vec{r}\,'\right)\right\}\sim\mathcal{O}\left(\left|\bm{m}\right|\mathcal{V}_{\vec{r}}/q\hbar\right)\simeq0$. The Lagrangian formalism allows us to introduce dissipation via a Rayleigh function.
\end{thebibliography}
\end{document}